\documentclass[11pt]{article}
\usepackage{amsbsy,amsfonts}
\DeclareMathAlphabet{\bi}{OML}{cmm}{b}{i}
\newcommand{\be}{\begin{equation}}
\newcommand{\ee}{\end{equation}}
\newcommand{\ba}{\begin{array}}
\newcommand{\ea}{\end{array}}

\newcommand{\ds}{\displaystyle}

\newtheorem{prop}{Proposition}

\begin{document}

\title{\protect\vspace*{-15mm}\bf Zero curvature representation\\ for a new fifth-order integrable system}

\author{A. Sergyeyev\\
Silesian University in Opava, Mathematical Institute,\\ Na
Rybn\'\i{}\v{c}ku 1, 746\,01 Opava, Czech Republic,\\
E-mail: \tt{Artur.Sergyeyev@math.slu.cz}}
\date{}
\maketitle
\begin{abstract}
In this brief note we present a zero-curvature representation
for one of the new integrable systems found by Mikhailov, Novikov and Wang
in nlin.SI/0601046.
\end{abstract}

\section*{Introduction}

The zero curvature representations (ZCRs) are well known to be of paramount importance
in the theory of integrable systems. Indeed, knowing a ZCR involving
an essential (spectral) parameter enables one to use
the inverse scattering transform (and/or find and use Darboux and B\"acklund transformations)
and construct plenty of exact solutions
for the system in question, including multisoliton and finite-gap solutions,
see e.g.\ the classical monographs \cite{as, tf, ne, ts} and references therein.

However, for integrable systems discovered
using the symmetry approach, see e.g.\
\cite{miknov, mnw, mss, msy, mikyam} and references therein,
there seems to be no straightforward procedure leading to
a ZCR suitable for the inverse scattering transform
(the ZCRs arising from the recursion operators often lead to pseudodifferential
spectral problems that are difficult to handle), so finding
``good" ZCRs for such systems
can be rather challenging.

In this paper we present a ZCR
for a new integrable system of the form
\be\label{e}
\ba{l}
u_t = u_{xxxxx}-20 u u_{xxx} -50 u_x u_{xx}+80 u^2 u_x-w_x,\\
w_t = -6 w u_{xxx}-2 u_{xx} w_x+96 w u u_x+16 w_x u^2
\ea
\ee
found by Mikhailov, Novikov and Wang \cite{mnw} using the symmetry approach.
This system is bihamiltonian and possesses a recursion operator \cite{mnw}.
More precisely, Eq.(\ref{e}) can
be obtained  \cite{mnw} from another integrable system, Eq.(101) of \cite{mnw},
through a simple invertible change
of variables, and hence the recursion operator, Hamiltonian and symplectic structure
found for Eq.(101) in \cite{mnw} can be readily transferred to (\ref{e}).

Setting $w=0$
reduces \cite{mnw} Eq.(\ref{e}) to the well-known Kaup--Kupershmidt equation
\be\label{kk}
u_t = u_{xxxxx}-20 u u_{xxx} -50 u_x u_{xx}+80 u^2 u_x.
\ee
The ZCR for (\ref{kk}) is readily extracted from the corresponding Lax pair
\cite{kaup} and reads
\be\label{zcrgen}
A_t-B_x+[A,B]=0,
\ee
where
\begin{equation}\label{zcr0}
\ba{l}
A=\left(\begin{array}{@{}ccc@{}} 0 & -2 & 0 \\ -u & 0 & -2 \\ \lambda &
-u & 0 \end{array}\right),\\[10mm]
B=\left(\begin{array}{@{}ccc@{}} 2 u_{xxx} - 32 u u_x + 24 \lambda u &
4 u_{xx} - 32 u^2 & -144 \lambda \\[3mm]
-u_{xxxx} + 18 u u_{xx} + 16 u_x^2 - 12 \lambda u_x - 16 u^3 + 72 \lambda^2 &
-48 \lambda u & 4 u_{xx} - 32 u^2 \\[3mm]
4\lambda u_{xx} - 20 \lambda u^2 &
b_{32} &
-2 u_{xxx} + 32 u u_x + 24 \lambda u \end{array}\right),
\ea
\end{equation}
and
\[
b_{32}=-u_{xxxx} + 18 u u_{xx} + 16 u_x^2 + 12 \lambda u_x - 16 u^3 + 72 \lambda^2.
\]

We succeeded in modifying the above $A$ and $B$ in such a way
that the resulting matrices go into (\ref{zcr0}) if $w=0$
{\em and} yield a ZCR for (\ref{e}) if $w\neq 0$:

\begin{prop}
The matrices \begin{equation}\label{zcr}
\ba{l}
A=\left(\begin{array}{@{}ccc@{}} 0 & -2 & 0 \\ -u & 0 & -2 \\
\ds\frac{1}{72} \frac{w}{\lambda} + \lambda & -u & 0 \end{array}\right),\\[9mm]
B=\left(\begin{array}{@{}ccc@{}} 2 u_{xxx} - 32 u u_x + 24 \lambda u &
4 u_{xx} - 32 u^2 & -144 \lambda \\[3mm]
-u_{xxxx} + 18 u u_{xx} + 16 u_x^2 - 12 \lambda u_x - 16 u^3 + 72 \lambda^2  + w&
-48 \lambda u & 4 u_{xx} - 32 u^2 \\[3mm]
4\lambda u_{xx} - 20 \lambda u^2-\ds\frac{w (u_{xx} - 8 u^2)}{36\lambda} &
b_{32} &
-2 u_{xxx} + 32 u u_x + 24 \lambda u \end{array}\right),
\ea
\end{equation} where
\[
b_{32}=-u_{xxxx} + 18 u u_{xx} + 16 u_x^2 + 12 \lambda u_x - 16 u^3 + 72 \lambda^2+w,
\]
provide a ZCR for (\ref{e}) with a spectral parameter $\lambda$, that is, the matrix equation
\[
A_t-B_x+[A,B]=0
\]
is equivalent to (\ref{e}) for all $\lambda\neq 0$.
\end{prop}

\section*{Acknowledgements}

This research was supported in part by
the Czech Grant Agency (GA\v CR) under grant No.\ 201/04/0538,
by the Ministry of Education, Youth and Sports of Czech Republic
under grant MSM 4781305904.
\looseness=-1


\end{document}